\documentclass[aps,pre,groupaddress,showpacs,onecolumn]{revtex4-1}
\usepackage{amsfonts}
\usepackage{bm}
\usepackage{epsfig,amsmath,graphicx,amssymb,overpic}
\usepackage{color}

\def\be{\begin{equation}}
\def\ee{\end{equation}}
\def\bee{\begin{eqnarray}}
\def\ene{\end{eqnarray}}
\def\bes{\begin{subequations}}
\def\ees{\end{subequations}}
\DeclareMathOperator{\e}{e}
\DeclareMathOperator{\sech}{sech}

\newcommand{\PT}{{\cal PT}}

\begin{document}

\title{Solitons and their stability in the nonlocal nonlinear Schr\"odinger equation \\ with $\PT$-symmetric potentials}
\author{Zichao Wen}
\author{Zhenya Yan}
\email{zyyan@mmrc.iss.ac.cn}
\affiliation{\vspace{0.1in}Key Laboratory of Mathematics Mechanization, Institute of Systems
Science, AMSS, Chinese Academy of Sciences, Beijing 100190, China \\
School of Mathematical Sciences, University of Chinese Academy of
Sciences, Beijing 100049, China}


\begin{abstract}

\centerline{ {\bf Chaos} 27 (2017) 053105.}
 
\vspace{0.2in} \baselineskip=15pt
We report localized nonlinear modes of the self-focusing and defocusing nonlocal nonlinear Schr\"odinger equation with the generalized $\PT$-symmetric Scarf-II, Rosen-Morse, and periodic potentials. Parameter regions are presented for broken and unbroken $\PT$-symmetric phases of linear bounded states, and the linear stability of the obtained solitons. Moreover, we numerically explore the dynamical behaviors of solitons and find stable solitons for some given parameters.

\vspace{0.3in}

\textbf{In recent years, linear and nonlinear partial differential equations having $\PT$ symmetry structures draw intense interests, because many significant and interesting properties in $\PT$-symmetric systems are absent in conventional Hermitian ones. Particularly, stable solitons can be generated in local nonlinear wave equations with $\PT$-symmetric external potentials. Recently, the nonlocal nonlinear Schr\"odinger (NNLS) equation has been presented and shown to be completely integrable, and it admits new properties that differ from its local version (the conventional nonlinear Schr\"odinger equation). Since the NNLS equation has a $\PT$-symmetric self-induced potential, it is straightforward to consider the NNLS equation in the presence of $\PT$-symmetric external potentials. However, to the best of our knowledge, there is no report on soliton solutions and their stability in this generalized nonlocal model. In this paper, we derive some exact localized nonlinear modes of the NNLS equation with three kinds of generalized $\PT$-symmetric potentials, and conclude graphically the relationship between model parameters and broken and unbroken $\PT$-symmetric phases of the generalized $\PT$-symmetric potentials. Moreover, we study linear stability and dynamical behaviors of the obtained solitons under the scope of the above-mentioned parameters.}

\end{abstract}
\maketitle
\baselineskip=15pt

\section{Introduction}

The nonlinear Schr\"odinger (NLS) equation is a fundamental model in many fields of nonlinear science~\cite{op,ocean} (alias the Gross-Pitaevskii equation in Bose-Einstein condensates~\cite{bec}). The NLS equation with different kinds of real-valued external potentials, such as harmonic, periodic and double-well potentials, have been intensely studied, including exact soliton solutions (see, e.g., ~\cite{hm, yan10} and references therein) and numerical soliton solutions (see, e.g., ~\cite{op, bec, RealPotential1, RealPotential2, RealPotential3} and references therein).

Motivated by the famous work of Bender and Boettcher~\cite{Bender98} and other related works~\cite{ms, hyper,sh,PP,PTR} that the complex $\PT$-symmetric potentials were introduced into the usual Hermitian Hamiltonians, Musslimani and his collaborators~\cite{Muss} introduced
the complex $\PT$-symmetric potentials (e.g., complex $\PT$-symmetric Scarf-II and periodic potentials) into the usual NLS equation such that the stable nonlinear modes were found. The imaginary parts of the complex $\PT$-symmetric potentials have an effect of gain-and-loss on nonlinear modes, and may lead to stable nonlinear modes since the gain-and-loss distributions with the $\PT$ symmetry can always be balanced. Here the linear parity operator $\mathcal{P}$ and antilinear time-reversal operator $\mathcal{T}$ are defined as $\mathcal{P}: x\rightarrow -x, \, p\rightarrow -p$, and $\mathcal{T}: x\rightarrow x,\, p\rightarrow -p,\, i\rightarrow -i$ defined~\cite{Bender2}.

After that, the NLS equation with distinct types of complex $\PT$-symmetric potentials have been studied to yield stable localized nonlinear modes~\cite{Mussg}. Also, the stable localized nonlinear modes are found for some generalized NLS equations in the presence of $\PT$-symmetric potentials such as the NLS equation with the momentum term~\cite{yanchaos15}, the third-order NLS equation~\cite{yansr}, the derivative NLS equation~\cite{yanpre17}, and etc. (see the recent review in ~\cite{vvk16}).

Recently, a new integrable nonlocal nonlinear Schr\"odinger (NNLS) equation introduced from the AKNS hierarchy is of the form~\cite{nnls}
\bee \label{nnls}
 i\,\frac{\partial}{\partial t}\psi(x,t) = -\frac{\partial^2}{\partial x^2}\psi(x,t) +g\,\psi^2(x,t)\psi^{*}(-x,t),
\ene
where $\psi(x,t)$ is a complex field, $\psi(-x,t)$ a nonlocal field, $g$ a non-zero real constant, and the star stands for the complex conjugate. The NNLS equation is shown to possess a Lax pair and infinite numbers of conservation laws~\cite{nnls}. Moreover, the higher-order rational solitons and dynamics of Eq.~(\ref{nnls}) with the defocusing case $(g=1)$ have been found by means of the generalized Darboux transformation methods and numerical methods~\cite{wen2016, yz}. Eq.~(\ref{nnls}) becomes the usual NLS equation if $\psi(x,t)$ is an even function about space. More recently, two families of two-parameter and multi-component extensions of Eq.~(\ref{nnls}) were also found~\cite{yanaml}.  To the best of our knowledge, localized nonlinear modes of the NNLS equation with any complex $\PT$-symmetric potential and their stability have not been investigated yet.

The rest of this paper is organized as follow. In section II, we present a general theory for the self-focusing and defocusing NNLS equation with $\PT$-symmetric potentials, linear broken and unbroken $\PT$-symmetric phases and the linear stability for the localized nonlinear modes. In section III, we investigate in sequence the generalized $\mathcal{PT}$-symmetric Scarf-II, Rosen-Morse, and Rosen-Morse-II potentials. Unbroken and broken $\PT$-symmetric phases, localized nonlinear modes, and their linear stability and dynamical behaviors are discussed in details.

\section{$\PT$-symmetric nonlocal nonlinear model and general theory}

We aim to investigate the NNLS equation with $\mathcal{PT}$-symmetric potentials
\bee
 i\,\frac{\partial}{\partial t}\psi(x,t)=-\frac{\partial^2}{\partial x^2}\psi(x,t) +[V(x)+i\,W(x)]\psi(x,t)+g\,\psi^2(x,t)\psi^{*}(-x,t),
\label{nls}
\ene
where $\psi(x,t)$ is a complex function of real variables $x$ and $t$, $\psi^{*}(-x,t)$ stands for the complex conjugate of the nonlocal field $\psi(-x,t)$, the constant $g$ describes two-body `self-focusing' ($g=-1$) or `defocusing' ($g=1$) interactions. The $\mathcal{PT}$-symmetric potential is required that $V(x)$ is an even function, i.e., $V(x)=V(-x)$, and $W(x)$ is an odd function, i.e., $W(x)=-W(-x)$. Eq.~(\ref{nls}) without $\PT$-symmetric potentials reduces to the NNLS equation introduced by Ablowitz and Musslimani, i.e., Eq.~(\ref{nnls}). Particularly, if the obtained solutions of Eq.~(\ref{nls}) have even parity symmetry for space, i.e., $\psi(-x,t)=\psi(x,t)$, they also solve the conventional NLS equation with the same $\PT$-symmetric potentials. On the other hand, if the obtained solutions of Eq.~(\ref{nls}) have no even parity symmetry for space, i.e., $\psi(-x,t)\neq\psi(x,t)$, they are not identical to the solutions of the conventional NLS equation. In general, Eq.~(\ref{nls}) with non-zero potentials (i.e., $V(x)+i\,W(x)\not\equiv 0) $ is not completely integrable. Let $Q(t)=\int^{+\infty}_{-\infty}\psi(x,t)\psi^{*}(-x, t)dx$ and $P(t)=\int^{+\infty}_{-\infty}|\psi(x,t)|^2dx$, named ``quasi-power" and ``power" respectively in the context of $\PT$-symmetric optics~\cite{Muss}, and it is easy to show that $dQ(t)/dt=0$, thus $Q(t)$ is a conserved quantity, and that $dP(t)/dt=\int^{+\infty}_{-\infty}|\psi(x,t)|^2\{2W(x)+g\,{\rm Im}[\psi(x,t)\psi^{*}(-x,t)-\psi(-x,t)\psi^{*}(x,t)]\}dx$, thus $Q(t)$ may not be conserved.

We focus on stationary solutions of the $\mathcal{PT}$-NNLS equation in the form $\psi(x,t)=\phi(x)\e^{-i\mu t}$, where $\mu$ is the propagation constant in optics or real chemical potential in BEC, and complex nonlinear eigenmode $\phi(x)$ satisfies the stationary $\PT$-NNLS equation
\bee\label{ode}
\mu\,\phi(x) = -\frac{d^2}{dx^2}\phi(x)+ \left[V(x)+i\,W(x)\right]\phi(x) +g\,\phi^2(x)\phi^{*}(-x),
\label{snls}\ene
subject to the boundary conditions $\phi(x)\to0$ as $x\to\pm\infty$. The linear problem of Eq.~(\ref{snls}) is written as
$H\,\Phi(x)=\lambda\,\Phi(x)$, where the Hamiltonian $H\!=\!-\partial^2_x +V(x)+i\,W(x)$ is a linear Schr\"{o}dinger operator with complex $\PT$-symmetric potential, and $\Phi(x)$ is the eigenfunction corresponding to eigenvalue $\lambda$. Usually, $H$ is parameterized by tuning parameter(s) in specific $V(x)$ and $W(x)$. In the interested parametric space, it is called the parametric region of {\it unbroken} $\PT$ symmetry
(or  unbroken $\PT$-symmetric phase), if all of the eigenvalues of the Hamiltonian are real in corresponding parametric region; otherwise the parametric region of {\it broken} $\PT$ symmetry (or broken $\PT$-symmetric phase).

When $W(x)\not=0$, the non-zero solution $\phi(x)$ should be complex, and thus can be written as
\bee \label{solub}
 \phi(x)=\hat\phi(x) \e^{i\varphi(x)},
\ene
where the amplitude $\hat\phi(x)$ is real and strictly positive, the real function $\varphi(x)$ denotes the phase. We substitute Eq.~(\ref{solub}) into Eq.~(\ref{ode}) and yield the relations between the amplitude and the phase
\bee\label{ode1}
\frac{[\varphi_x(x)\hat\phi^2(x)]_x}{\hat\phi^{2}(x)} = W(x)+g\,\hat\phi(x)\hat\phi(-x)\sin[\theta(x)],
\ene
and
\bee \label{ode2}
\frac{\hat\phi_{xx}(x)}{\hat\phi(x)}=\mu+V(x)+g\,\hat\phi(x)\hat\phi(-x)\cos[\theta(x)],
\ene
where $\theta(x)=\varphi(x)-\varphi(-x)$, which differs from the local NLS cases~\cite{Muss, Mussg}.

For given $\PT$-symmetric potential $V(x)+i\,W(x)$, one can find the exact solutions by solving Eqs.~(\ref{ode1}) and (\ref{ode2}), or numerical solutions via applicable numerical methods in principle. Further, one can study the linear stability of the obtained localized modes by considering the perturbed solution of $\mathcal{PT}$-NNLS equation (\ref{nls}) in the form
\begin{eqnarray}\label{perturbation}
\psi(x,t) = \left\{ \phi(x) + \varepsilon\left[F(x)\e^{-i\delta t}+G^*(-x)\e^{i\delta^*t}\right]\right\}\e^{-i\mu t},
\end{eqnarray}
where $\varepsilon\ll1$, $F(x)$ and $G(x)$ are the perturbation eigenfunctions. Via the substitution of Eq.~(\ref{perturbation}) into Eq.~(\ref{nls}) and the linearization with respect to $\varepsilon$, the linear eigenvalue problem for the perturbation eigenfunctions is given by
\begin{eqnarray}
\left[\begin{array}{cc}    L(x) & g\,\phi^2(x) \vspace{0.1in} \\   -g\,\phi^{*2}(-x) & - L(x)   \end{array}\right]
\left[  \begin{array}{c}    F(x) \vspace{0.1in}\\    G(x)   \end{array} \right]
  =\delta \left[  \begin{array}{c}    F(x) \vspace{0.1in}\\    G(x)   \end{array}\right],
\end{eqnarray}
where $ L(x)=-\partial^2_x+V(x)+i\,W(x)+2 g\,\phi(x)\phi^{*}(-x)-\mu$ is $\mathcal{PT}$-symmetric, i.e., $ L(x)= L^{*}(-x)$. It is the routine that the $\PT$-symmetric nonlinear modes are linearly stable if all eigenvalues $\delta$ of this problem are real, otherwise they are linearly unstable.

In the following, several interesting and physically relevant $\mathcal{PT}$-symmetric potentials are introduced in Eq.~(\ref{nls}) and the properties of corresponding nonlinear modes are to be discussed.

\section{Nonlinear modes with $\PT$-symmetric potentials}

\subsection{Generalized Scarf-II potential}

We first consider the generalized $\mathcal{PT}$-symmetric complex Scarf-II potential $V_1(x)+i\,W_1(x)$, with the components
\bee \label{poten}
\left[\!\!\begin{array}{cc}   V_1(x) \vspace{0.05in}\\  W_1(x) \end{array}\!\!\right]
\!=\!-\!\left[\!\!\begin{array}{cc}   (w_1^2+2)\,\sech^2(x) \vspace{0.05in}\\  3w_1\sech (x)\tanh(x) \end{array}\!\!\right]
\!\!-\!\sigma_1(x)\!\!\left[\!\!\begin{array}{cc} \cos[\theta_1(x)] \vspace{0.05in}\\  \sin[\theta_1(x)]\end{array}\!\!\right], \quad
\ene
where
\bee
\sigma_1(x)=g\rho_1^2\sech^2(x),\quad \theta_1(x)=2w_1\tan^{-1}[\sinh(x)],
\ene
and $w_1,\, \rho_1$ are real-valued constants.

The linear eigenvalue problem for the $\PT$-symmetric Scarf-II potential (\ref{poten}) related to Eq.~(\ref{nls}) as
\bee \label{ls1}
 H_1\,\Phi(x)=\lambda\,\Phi(x),\quad H_1=-\partial_x^2 + V_1(x)+ i\,W_1(x),
\ene
where $\lambda$ and $\Phi(x)$ are the eigenvalue and eigenfunction, respectively, and $\Phi(x)=0$ as $x\to\pm\infty$. Notice that the generalized Scarf-II potential (\ref{poten}) reduces to the conventional $\PT$-symmetric complex Scarf-II potential when $\rho_1=0$, whose linear problem can be shown to admit an entirely real spectrum for any $w_1$, since $3|w_1|\leq 9/4+w_1^2$ always holds~\cite{sh}.

\begin{figure}[htb]
	\begin{center}
      {\scalebox{0.6}{\includegraphics{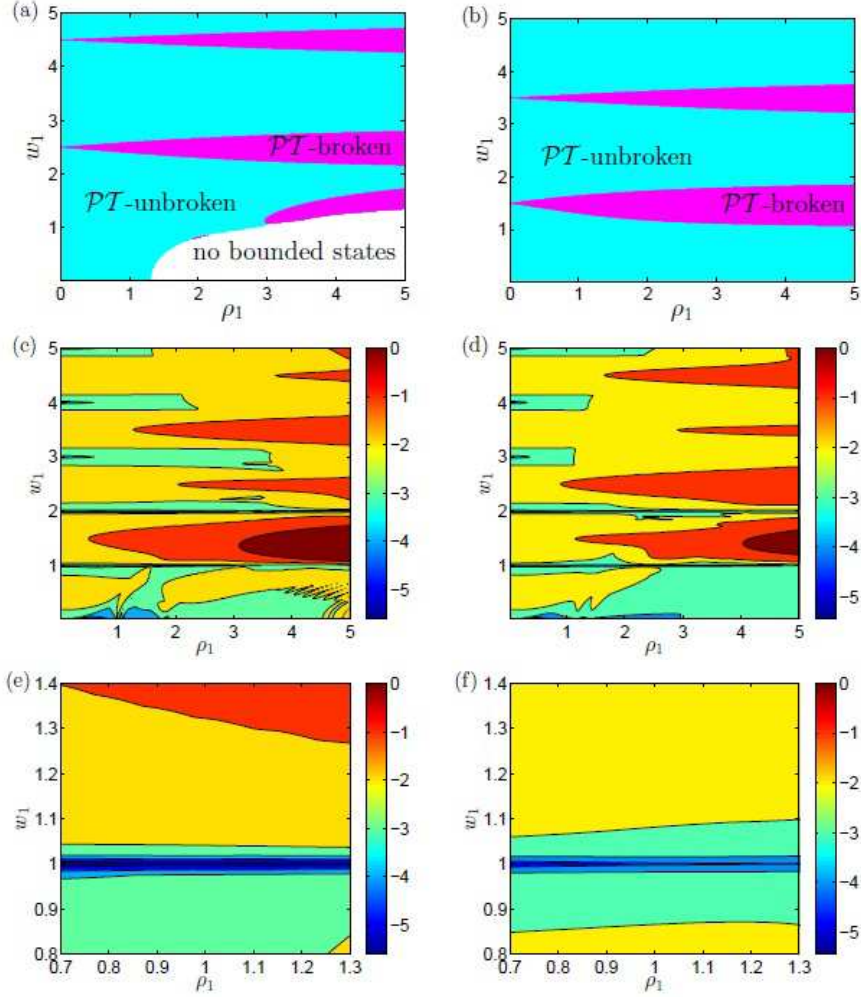}}}
	\end{center}
	\vspace{-0.25in} \caption{\small (a, b) parametric regions ($\rho_1,\, w_1$) of broken and unbroken $\PT$ symmetry of linear bounded states for the generalized $\PT$-symmetric Scarf-II potential; (c)-(f) maximal absolute values of imaginary parts of linearization eigenvalues $\delta$ in parametric regions ($\rho_1,\, w_1$) (common logarithmic scale), i.e.,  $\max\left\{\log|\text{Im}[\sigma_p(L_1)]|\right\}$. In (a) and (b), cyan means the $\PT$-unbroken regions, purple the $\PT$-broken ones, and blank means no bounded states at all. The left column is with $g=-1$, and the right one with $g=1$. } \label{fig-1}
\end{figure}

\begin{figure}[htb]
	\begin{center}
    {\scalebox{0.6}{\includegraphics{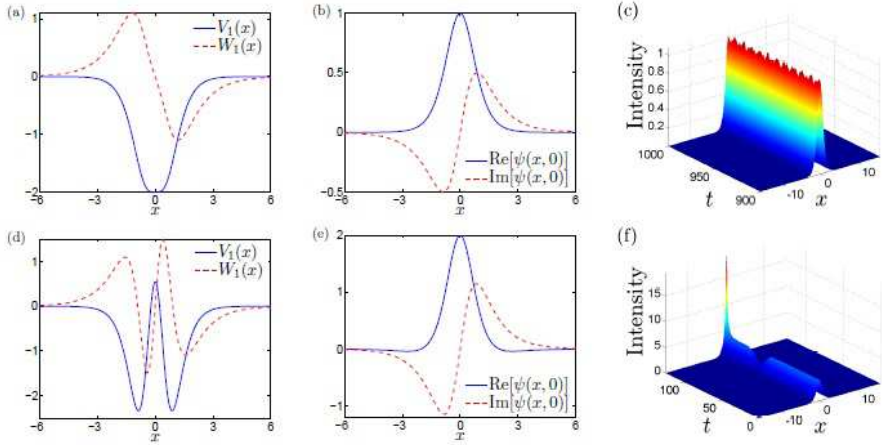}}}
	\end{center}
	\vspace{-0.15in}\caption{\small Profiles of (a) single-well potential and (d) double-well potential, (b,e) real and imaginary parts of initial exact nonlinear modes (\ref{solu}), and the wave propagations for the modes (\ref{solu}) for (c) stable oscillation with $2\%$ initial random noise, and (f) unstable case which diverges around $t=100$ without an initial noise. For (a, b, c), the parameters are $w_1=1, \rho_1=1$; for (d, e, f) the parameters are $w_1=1.2, \rho_1=2$. All the cases are in unbroken $\PT$-symmetric phase and of the self-focusing case $g=-1$.} \label{fig-2}
\end{figure}


Here we consider the linear problem (\ref{ls1}) associated with the generalized Scarf-II $\PT$-symmetric potential (\ref{poten}) at $\rho_1\not=0$. We fix the self-focusing ($g=-1$) or defocusing ($g=1$) nonlinearities and investigate the parametric regions $(w_1,\rho_1)$ of broken and unbroken $\PT$ symmetry (see Fig.~\ref{fig-1}).
The generalized Scarf-II $\PT$-symmetric potential (\ref{poten}) is a combination of hyperbolic and periodic functions, and the periodic parts, which are parameterized by $w_1$, result in alternate patterns of broken and unbroken $\PT$-symmetric phases with respect to $w_1$, which differs from the conventional Scarf-II $\PT$-symmetric potential. In the self-focusing case ($g=-1$), the real part of $\PT$-symmetric potential becomes shallower when $\rho_1$ increases, which makes it harder for bounded states to come into being, and that results in the bottom-right region of no bounded states (see Fig.~\ref{fig-1}(a)). However, the defocusing case ($g=1$) is in the opposite and therefore bounded states exist when $\rho_1\geqslant0$, at least in the region we have searched as shown in Fig.~\ref{fig-1}(b). It should be noted that the pictures Figs.~\ref{fig-1}(a) and (b) are symmetric with respect to the horizontal axis $w_1=0$ because of the odd symmetry of $w_1$ in the linear eigenvalue problem.

For the given $\mathcal{PT}$-symmetric potential (\ref{poten}), we can find exact bright soliton solutions of $\mathcal{PT}$-NNLS equation (\ref{snls})
\bee\label{solu}
 \phi_1(x)=\rho_1\sech(x)\e^{iw_1\tan^{-1}[\sinh(x)]},
\ene
with $\mu=-1$. It follows from the solution (\ref{solu}) that we have
\bee
 S_1(x)=\frac{i}{2}(\phi\phi^*_x-\phi_x\phi^*)=\rho_1^2w_1\sech^3(x).
  \ene
 Following the idea in the $\PT$-symmetric classical optics~\cite{Muss}, we know that $S_1(x)$ is everywhere positive in the $\PT$ cell and the power always flows in one direction when the positive $w_1$, i.e., from the gain toward the loss domain. We can see in Figs.~\ref{fig-1}(c)-(f) that, for the exact soliton solusions (\ref{solu}), the parametric region for linear stable solitons is tiny in the whole parametric space.

We next investigate the dynamical stability of nonlinear modes (\ref{solu}) for both self-focusing and defocusing cases by numerical simulations for the wave propagation without or with an initial random perturbation of order $2\%$.

For the self-focusing case ($g=-1$), Fig.~\ref{fig-2} illustrates the profiles of potentials and exact initial soliton states, and numerical simulations for the wave propagations. Interesting enough, stability of the solitons is sensitive to the shape of the real potential such as single-well or double-well, and thus in the following we focus on the dependence of soliton stability on the shape of potentials.
When $w_1=\rho_1=1$ corresponding to a linear stable case, it locates at a $\PT$-unbroken region for the linear operator $H_1$ [cf. Eq.~(\ref{ls})] in the parametric space. The corresponding nonlinear mode is stable and an evident oscillatory (breather-like) behavior can be observed (see Fig.~\ref{fig-2}(c)). If we choose $w_1=1.2,\, \rho_1=2$ corresponding to a linear unstable case, the linear operator $H_1$ is $\PT$-unbroken again, and the double-well potential $V_1(x)$ has two completely separated wells (see Fig.~\ref{fig-2}(d)). In this case the nonlinear mode diverge at around $t=100$ without an initial noise (see Fig.~\ref{fig-2}(f)).


\begin{figure}[htb]
    \begin{center}
        {\scalebox{0.6}{\includegraphics{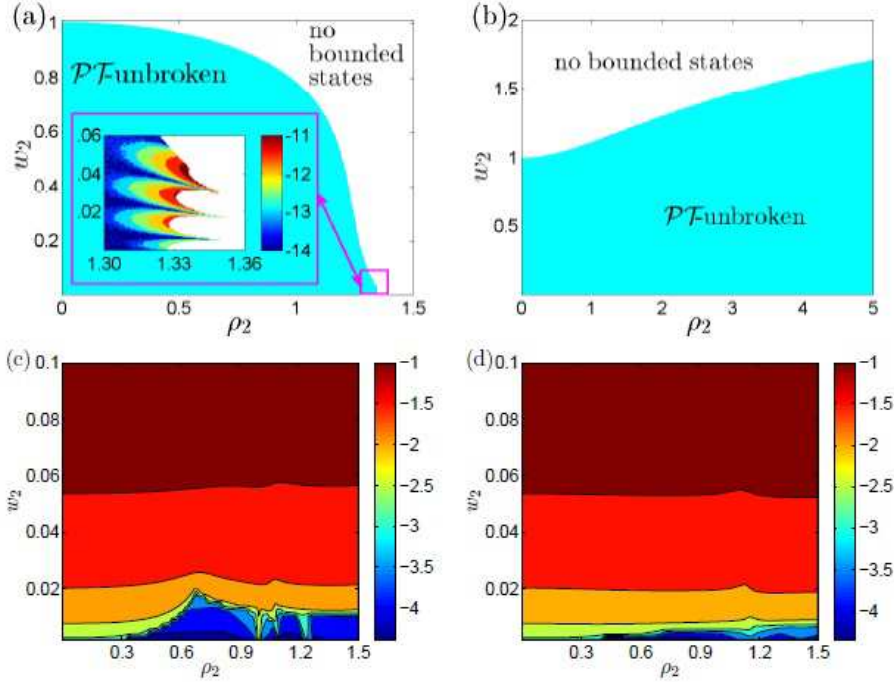}}}
	\end{center}
	\vspace{-0.15in}\caption{\small (a, b) parametric regions ($\rho_2,\, w_2$) of broken and unbroken $\PT$ symmetry of linear bounded states for the generalized $\PT$-symmetric Rosen-Morse potential; (c, d) maximal absolute values of imaginary parts of linearization eigenvalues $\delta$ in parametric regions ($\rho_1,\, w_1$) (common logarithmic scale), i.e., $\max\left\{\log_{10}|\text{Im}[\sigma_p(L_2)]|\right\}$. In (a) and (b) cyan means the $\PT$-unbroken region and blank means no bounded states at all. Detailed picture inside (a) displays the max absolute values of logarithm imaginary parts of the whole discrete spectra, i.e., $\max\left\{\log|\text{Im}[\sigma_p(H_2)]|\right\}$, within the parametric region $1.3<\rho_2<1.36$ and $0<w_2<0.06$.} \label{fig-4}
\end{figure}


\begin{figure}[htb]
	\begin{center}
    {\scalebox{0.6}{\includegraphics{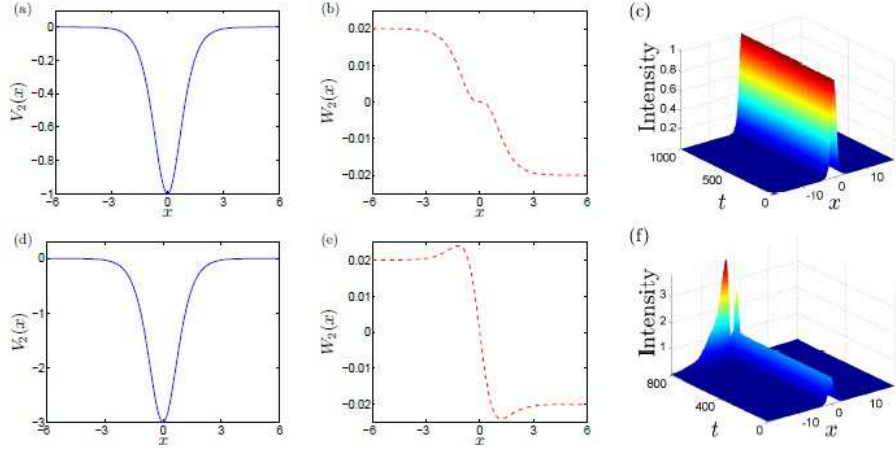}}}
	\end{center}
	\vspace{-0.15in}\caption{\small  (a,d) real and (b, e) imaginary parts of the generalized $\PT$-symmetric Rosen-Morse potentials (\ref{poten-2}), and (c, f) the wave propagations of exact modes (\ref{solu2}) with $2\%$ initial random noise. (a, b, c) corresponding to the self-focusing case $g=-1$; (d, e, f) corresponding to the defocusing case $g=1$. Notice that the corresponding propagation is not diverge within $t=1000$ if we add no initial noise in this case, which is not shown here. Parameters are $w_2=0.01, \rho_2=1$ in both cases.} \label{fig-5}
\end{figure}

\subsection{Generalized Rosen-Morse potential well}

In this subsection, we study the generalized $\PT$-symmetric Rosen-Morse potential
\bee \label{poten-2}
\left[\!\!\begin{array}{cc}   V_2(x) \vspace{0.05in}\\  W_2(x) \end{array}\!\!\right]
\!=\!-\!\left[\!\!\begin{array}{cc}   2\,\sech^2(x) \vspace{0.05in}\\  2w_2\tanh(x) \end{array}\!\!\right]
\!\!-\!\sigma_2(x)\!\!\left[\!\!\begin{array}{cc} \cos[\theta_2(x)] \vspace{0.05in}\\  \sin[\theta_2(x)] \end{array}\!\!\right],
\ene
with \bee
\sigma_2(x)=g\rho_2^2\sech^2(x),\quad \theta_2(x)=2w_2x,
\ene and $w_2,\, \rho_2$ being real-valued constants. When
$\rho_2=0$, the $\PT$-symmetric potential $V_2(x)+i\,W_2(x)$ becomes the conventional $\PT$-symmetric Rosen-Morse potential~\cite{Levai00}.

The linear eigenvalue problem for the $\PT$-symmetric Rosen-Morse potential (\ref{poten-2}) related to Eq.~(\ref{nls}) is
\bee \label{ls}
 H_2\,\Phi(x)=\lambda\,\Phi(x),\quad H_2=-\partial_x^2+\! V_2(x)+ i\,W_2(x),
\ene
where $\lambda$ and $\Phi(x)$ are the eigenvalue and eigenfunction, respectively, and $\Phi(x)\to0$ as $x\to\pm\infty$. The parametric regions for $(w_2,\rho_2)$ of broken and unbroken $\PT$ symmetry are shown in Figs.~\ref{fig-4} (a) and (b).

For the given $\PT$-symmetric Rosen-Morse potential (\ref{poten-2}), we can find soliton solutions of the $\mathcal{PT}$-NNLS equation (\ref{snls}) in the form
\bee\label{solu2}
 \phi_2(x)=\rho_2\sech(x)\e^{i w_2 x}
\ene
with $\mu=w_2^2-1$, and $S_2(x)=i\,(\phi\phi^*_x-\phi_x\phi^*)/2=\rho_2^2w_2\sech^2(x)$. $S_2(x)$ is everywhere positive in the $\PT$ cell for positive $w_2$, and the power always flows in one direction, i.e., from the gain toward the loss domain. We can conclude from Figs.~\ref{fig-4}(c) and (d) that, linear stable exact solitons can only be found when $w_2$ is very close to zero.

Dynamical stability of the soliton solutions in Eq.~(\ref{solu2}) is checked via numerical simulations for wave propagations without or with initial random perturbation of order $2\%$, and all the cases here have unbroken $\PT$ symmetry for the linear operator $H_2$ [cf. Eq.~(\ref{ls})]. If we choose $w_2=0.01,\, \rho_2=1$, we have a stable propagation (see Fig.~\ref{fig-5}(c)) with $2\%$ initial random noise for the self-focusing case $g=-1$, but an unstable one (see Fig.~\ref{fig-5}(f)) without initial noise for the defocusing case $g=1$. Real and imaginary parts of the generalized $\PT$-symmetric Rosen-Morse potentials for the cases above are also shown in Fig.~\ref{fig-5}.

\subsection{Generalized Rosen-Morse-II (periodic) potential}

\begin{figure}[htb]
	\begin{center}
    {\scalebox{0.6}{\includegraphics{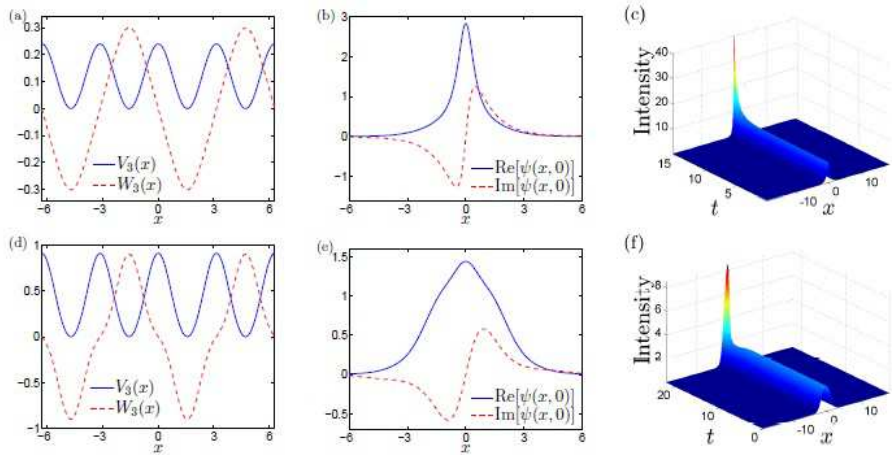}}}
	\end{center}
	\vspace{-0.15in}\caption{\small Profiles of the generalized $\PT$-symmetric Rosen-Morse-II potentials (\ref{poten-3}) (left column),
 numerical bright soliton solutions (middle column), and unstable wave propagations (right column).
  (a, b, c) $w_3=0.1,\, \rho_3=0.5,\, \mu=0.8,\, g=-1$ without an initial noise (the first row); (d, e, f) $w_3=0.3,\, \rho_3=1,\, \mu=0.4,\, g=-1$ without an initial noise (the second row).} \label{fig-6}
\end{figure}

We next consider another kind of $\PT$-symmetric potential which could be realized in optical lattice, the generalized Rosen-Morse-II potential:
\bee \label{poten-3}
\left[\!\!\begin{array}{cc}   V_3(x) \vspace{0.05in}\\  W_3(x) \end{array}\!\!\right]
\!=\!-\!\left[\!\!\begin{array}{cc}   w_3^2\cos^2(x) \vspace{0.05in}\\  3w_3\sin(x) \end{array}\!\!\right]
\!\!-\!\sigma_3(x)\!\!\left[\!\!\begin{array}{cc} \cos[\theta_3(x)] \vspace{0.05in}\\  \sin[\theta_3(x)]\end{array}\!\!\right],
\ene
with
\bee
\sigma_3(x)=g\rho_3^2\cos^2(x),\quad \theta_3(x)=2w_3\sin(x)
\ene
and $w_3,\, \rho_3$ being real-valued constants (see Figs.~\ref{fig-6}(a) and (b)).

For the given periodic potential (\ref{poten-3}), we can also find the periodic-wave solution of the $\mathcal{PT}$-NNLS equation (\ref{snls}) in the form
\bee\label{solu3}
 \phi_3(x)=\rho_3\cos(x)\e^{i w_3\sin(x)}
\ene
with $\mu=1$. Notice that under the chosen periodic potentials (\ref{poten-3}) even though $\hat{\phi}(x)=\rho_3\cos(x)$ is not positive everywhere,
Eq.~(\ref{solu3}) is still a solution of the $\mathcal{PT}$-NNLS equation (\ref{snls}). It follows from the solution (\ref{solu3}) that we have
\bee
S_3(x)=\frac{i}{2}(\phi\phi^*_x-\phi_x\phi^*)=\rho_3^2w_3\cos^3(x).
 \ene
It should be noted that for the positive $w_3$, $S_3(x)$ is no longer positive everywhere in the $\PT$ cell and the power does not always flows from the gain toward the loss domain.

By numerical methods, we obtain stationary soliton solutions of Eq.~(\ref{ode}) with the $\PT$-symmetric periodic potential (\ref{poten-3}) for the self-focusing ($g=-1$) case. Figs.~\ref{fig-6}(b) and (e) display the real and imaginary parts of the numerical bright solition solutions for the different parameters. Results for the numerical propagation are shown in Figs.~\ref{fig-6}(c) and (f), with $w_3=0.1, \rho_3=0.5, \mu=0.8$ and $w_3=0.3, \rho_3=1, \mu=0.4$, respectively. We find that the two numerical bright soliton solutions are both unstable in this case.

\section{Conclusions}

In conclusion, we have found localized nonlinear modes of the nonlocal nonlinear Schr\"odinger equation in the presence of generalized $\mathcal{PT}$-symmetric Scarf-II, Rosen-Morse, and Rosen-Morse-II potentials. We have investigated the parametric regions for the broken and unbroken $\PT$-symmetric phases. Moreover, we have studied the linear stability and dynamical stability of the obtained soliton solutions under the scope of the above-mentioned parameters. It should be note that all the obtained exact solutions above are exceptional ones, and one may find more generic localized solutions for the same potentials in a numerical form. The idea used in this paper can also be extended to other nonlocal nonlinear wave equations with $\PT$-symmetric potentials.

\acknowledgments

The authors thank the referees for their valuable comments and suggestions. This work was partially supported by the NSFC under Grant Nos. 11571346 and
61621003, and the Youth Innovation Promotion Association CAS.



\end{document}